\documentclass[%
 reprint,
 amsmath,amssymb,
 aps,
]{revtex4-2}

\usepackage{graphicx}
\usepackage{dcolumn}
\usepackage{bm}
\usepackage{lineno}
\usepackage{siunitx}
\usepackage{physics}

\begin{document}

\preprint{APS/123-QED}

\title{Angle dependence of $^{15}$N nuclear spin dynamics in diamond NV centers}

\author{Yusuke Azuma}
 \email{azuma.yusuke.sj@alumni.tsukuba.ac.jp}
\affiliation{
Division of Physics, Univ. of Tsukuba, 1-1-1 Tennodai, Tsukuba, Ibaraki, 305-8571, Japan
}

\author{Hideyuki Watanabe}%
\affiliation{
National Institute of Advanced Industrial Science and Technology (AIST) Central2, Umezono, Tsukuba, Ibaraki, 305-8568, Japan
}

\author{Satoshi Kashiwaya}%
\affiliation{
Department of Applied Physics, Nagoya Univ. Chikusa-Ku, Nagoya, Aichi, 464-8571, Japan
}

\author{Shintaro Nomura}
\affiliation{
Division of Physics, Univ. of Tsukuba, 1-1-1 Tennodai, Tsukuba, Ibaraki, 305-8571, Japan
}

\begin{abstract} 
We report on the dynamics of the Rabi oscillation and the Larmor precession of $^{15}N$ nuclear spin using nonselective short microwave pulses for initialization of $^{15}N$ nuclear spins.
We observe the Larmor precession of $^{15}N$ nuclear spin depending on the angle between the applied magnetic field and the axis of the nitrogen vacancy center. 
We propose to utilize the change of the Larmor frequency of the nuclear spins to detect static magnetic fields at high sensitivity.
Our results may contribute to enhancing the sensitivity of dc magnetic fields and devising novel protocols using $^{15}N$ nuclear spin in nitrogen vacancy centers in diamonds.
\end{abstract}

\maketitle
\section{Introduction}

High-sensitive measurements of magnetic fields are desired for applications in various fields, such as materials science and biomedicine. Quantum sensing is a method of measuring physical quantities using a quantum system and is attracting much attention for its potential to achieve higher sensitivity than conventional methods.\cite{degen2017quantum}
In particular, nitrogen-vacancy (NV) centers in diamonds \cite{Doherty_2013, Barry20} have attracted considerable attention as an outstanding system for quantum information processing at room temperature and atmospheric pressure.
The NV center in diamonds consists of two adjacent carbon atoms replaced by nitrogen and a vacancy.
Both electron spin and nuclear spins in diamond can be used for quantum information processing, and are controlled by microwave (MW) or radio-frequency waves (RF) pulses.
A nuclear spin is highly isolated from the environment, and can be used to store information as a quantum memory.\cite{Jiang09} An electron spins are used in sensing because they interact more strongly with the surrounding environment.

It has been demonstrated that a hybrid system of an electron and nuclear spins in diamonds enhances the sensitivity of magnetic field. ~\cite{Zaiser16, matsuzaki2016hybrid, rosskopf2017quantum, Pfender17ncomm}
The electron spin accumulates the phase from the magnetic field by the coupling to the external magnetic field, and the accumulated phase is transferred and stored in the nuclear spin.
This method effectively increases the sensitivity of the sensor by exploiting long coherence time of the nuclear spin.
An understanding of the hyperfine interaction in the electron state of the NV center~\cite{PhysRevB.79.075203} is essential to utilize the nuclear spins to enhance the sensitivity.

A drawback of utilizing nuclear spins to enhance the sensitivity is that the speed of nuclear spin state control is often slow, and initialization takes time. 
One approach is to use the interlevel anticrossing of the ground state or excited state of the NV center electron spin as a method to initialize the N nuclear spin of the diamond NV center. This method has drawbacks in that the bias magnetic field have to be fixed at $\approx$51 or $\approx$102 mT and that a small change in the bias magnetic field may affect the N nuclear spin polarization.~\cite{PhysRevB.47.8809, PhysRevLett.102.057403, PhysRevB.102.224101}
The other approach is to use a microwave pulse with a narrow linewidth to selectively excite hyperfine-split levels.\cite{rosskopf2017quantum} To this end, the duration of the microwave pulse $\tau_{\rm MW}$ has to be sufficiently long, typically, longer than $\tau_{\rm MW} \geq 1$ $\mu$s.
The N nuclear spin state is read out either optically~\cite{Jiang09,Zaiser16, matsuzaki2016hybrid, rosskopf2017quantum, Pfender17ncomm,PhysRevB.79.075203,PhysRevB.47.8809, PhysRevLett.102.057403, PhysRevB.102.224101,rosskopf2017quantum} or electrically~\cite{Morishita20, Gulka21} through the NV electron spin. 

We have recently proposed a method for initialization of $^{15}$N nuclear spins using nonselective microwave pulses to significantly reduce the time to control the nuclear spin state.~\cite{Azuma22}
This method enables us fast quantum control of nuclear spins.
By using this method, we report on the dynamics of the Rabi oscillation and the Larmor precession of $^{15}$N nuclear spin by varying the angle between the applied static magnetic field and the axis of the NV center.
We measure the Larmor frequency of the nuclear spin depending on the external transverse magnetic field, and compare with a model calculation that takes into account the hyperfine interactions.
Finally, we propse a high-sensitive static magnetic field sensing utilizing nuclear spins.

\section{Model calculation of effective magnetic rotation ratio of the nuclear spin}

\begin{figure}[htbp]
\includegraphics[width=85mm]{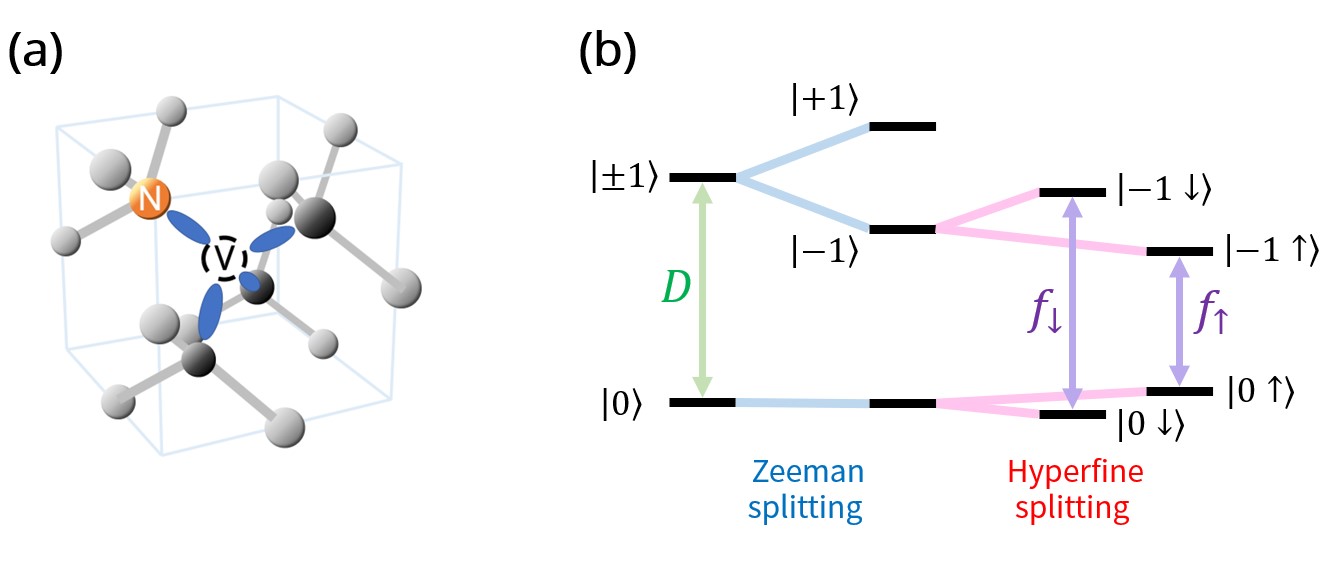}
\caption{NV center in diamond.
(a) Crystal structure of the NV center. (b) Energy levels in the ground state of the NV center. In a zero magnetic field, the energy at $\ket{m_S=\pm 1}$ is higher than at $\ket{0}$. In addition, an external magnetic field induces Zeeman splitting. The hyperfine interactions cause further splitting.}
  \label{fig:1}
\end{figure}

We study ensemble of NV centers formed in diamonds by implantation of $^{15}$N ions.~\cite{ofori2012spin,mariani2020system}
The implanted $^{15}$N is distinguished from naturally abundant $^{14}$N with an isotopic abundance ratio of 99.6\%, originally contained in the crystal as an impurity.
$^{15}$N has spin $I=1/2$ and takes on a simpler spin structure than $^{14}$N with $I=1$. 

Under an external magnetic field 
$\bf{B} =\qty(\it{B_x},\it{B_y},\it{B_z})$, 
the ground state Hamiltonian $\hat{H}$ of the $^{15}$NV centers can be written as~\cite{Maze08ElectronSpin,Doherty_2013}
\begin{align}
\label{eq:1}
\hat{H}/\hbar = D \hat{S}_z^2 + \gamma_e \vb*{B} \cdot \vb*{\hat{S}} -  \gamma_n \vb*{B} \cdot \vb*{\hat{I}} + \vb*{\hat{S}} \cdot \vb{A} \cdot \vb*{\hat{I}}.
\end{align}
where $\vb*{\hat{S}} = \qty(\hat{S}_x,\hat{S}_y,\hat{S}_z)$ and $\vb*{\hat{I}} = \qty(\hat{I}_x,\hat{I}_y,\hat{I}_z)$ are the electron spin and nuclear spin operators, respectively, with electron and nuclear gyromagnetic ratios $\gamma_e = 2\pi \times 28.0 \si{\,MHz/mT}$ and $\gamma_n = 2\pi \times (- 4.32) \si{\,kHz/mT}$.
The direction of the NV axis is set to the $z$ axis. The zero field splitting along the NV axis is $D = 2 \pi \times \SI{2.87}{GHz}$.
The hyperfine interaction is described by the diagonal tensor 
\begin{equation}
  \vb{A} = \mqty(A_{\perp}&0&0\\0&A_{\perp}&0\\0&0&A_{\parallel}),
\end{equation}
with transverse and longitudinal components $A_{\perp}= 2 \pi \times \SI{3.65}{MHz}$ and $A_{\parallel}= 2 \pi \times \SI{3.03}{MHz}$, respectively [Fig.\ref{fig:1}].

The angle between the $z$ axis and the direction of the external magnetic field $\vb*{B}$ is defined as $\theta$.
Expanding the Hamiltonian in Eq. (\ref{eq:1}), assuming that the magnetic field angle $\theta$ lies in the $x$-$z$ plane, we obtain
\begin{eqnarray}
\label{eq:2}
\hat{H}/\hbar = D \hat{S}_z^2 + \gamma_e \qty( B_x \hat{S}_x + B_z \hat{S}_z ) -  \gamma_n \qty( B_x \hat{I}_x + B_z \hat{I}_z ) \nonumber\\
+ A_{\parallel} \hat{S}_z \hat{I}_z + A_{\perp} \qty(\hat{S}_x \hat{I}_x + \hat{S}_y \hat{I}_y) .
\end{eqnarray}
Following the descriptions given in Refs. \cite{childress2006coherent} and \cite{oon2022ramsey}, we transform the Hamiltonian using a perturbation theory.
After the unitary transformations to the double-rotating electron and the rotating spin frames, the Hamiltonian $\hat{\tilde{H}}$ in the rotational coordinate frame becomes
\begin{eqnarray}
  \hat{\tilde{H}} = - \gamma_n \qty( \vb*{\beta}_{ind} + \vb*{\beta}(m_S) ) \cdot \vb*{\hat{I}},
\end{eqnarray}
where $\vb*{\beta}_{ind}$ is the magnetic field, independent of the electron spin, along the $z'$-axis as redefined for the rotating system,
\begin{align}
  \vb*{\beta}_{ind} = \beta_{ind} \hat{z'} = \sqrt{B^2_z + \qty( 1 + 2 \frac{\gamma_e A_{\perp}}{\gamma_n D})^2 B^2_x } \hat{z'},
\end{align}
and $\vb*{\beta}(m_S)$ is the magnetic field, which depends on the spin state of the electron,
\begin{equation}
  \vb*{\beta}(m_S) \nonumber\\
  =\frac{1}{\beta_{ind}} \mqty( B_x \qty{m_S \qty(1 + 2 \frac{\gamma_e A_{\perp}}{\gamma_n D}) \frac{A_{\parallel}}{\gamma_n} - 3 m^2_S \frac{\gamma_e A_{\perp}}{\gamma_n D} B_z} \\ 0 \\ - m_S \frac{A_{\parallel}}{\gamma_n} B_z - 3 m^2_S \frac{\gamma_e A_{\perp}}{\gamma_n D} \qty(1 + 2 \frac{\gamma_e A_{\perp}}{\gamma_n D}) B^2_x ).
\end{equation}

The effective nuclear Larmor frequency is then given by,
\begin{widetext}
\begin{equation}
  \omega_{m_S} = \abs{\gamma_n \qty( \vb*{\beta}_{ind} + \vb*{\beta}(m_S) ) } 
  =  - \gamma_n \sqrt{ B_x^2 \qty(1 + 2 \frac{\gamma_e A_{\perp}}{\gamma_n D} - 3 m_S^2 \frac{\gamma_e A_{\perp}}{\gamma_n D} )^2 + \qty(m_S \frac{A_{\parallel}}{\gamma_n} - B_z )^2 }.
  \label{eq:ms}
\end{equation}
\end{widetext}

\section{Experimental}

\begin{figure*}[htbp]
  \includegraphics[width=140mm]{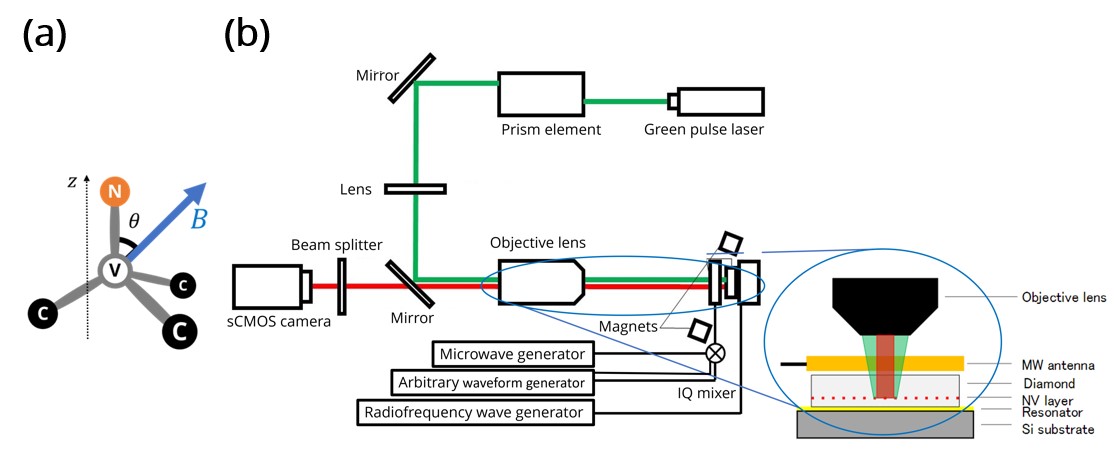}
  \caption{Experimental setup. (a) Relationship between the NV center axis and the magnetic field angle $\theta$. The direction of the NV axis is set to be parallel to the $z$-axis, The angle $\theta$ is the angle between the $z$-axis and the static magnetic field $B$. (b) Schematics of the measurement system. Laser pulses are incident to the (001) surface of the diamond and the emitted photoluminescence is collected by an objective lens and read by a scientific CMOS camera. Permanent magnets are mounted on a stage that can be rotated in two directions to change the direction of the external magnetic field.}
  \label{fig:2}
\end{figure*}

We used a (100)- oriented ultra-pure diamond chip (Element Six Ltd., electronic grade) with a size of $2.0 \times 2.0 \times \SI{0.5}{mm\cubed}$ \cite{mariani2020system,nomura2021near,ofori2012spin}.
After ion implantation of $^{15}$N$_2^+$, the diamond chip was annealed at $\SI{800}{\degreeCelsius}$. 
NV centers were created about $\SI{10}{nm}$ below the surface of the diamond chip.
Inhomogeneous dephasing time $T_{2}^{*}$ of the diamond chip was estimated to be 0.8 $\mu$s.

Figure \ref{fig:2}(b) shows the schematics of the experimental setup.
Magnetic field $\vb*{B}$ was applied by a pair of Nd$_2$Fe$_{14}$B permanent magnets. The external magnetic field was $\SI{4.0}{mT}$.
The photoluminescence from NV centers was imaged at room temperature with a wide-field microscope equipped with a cooled scientific CMOS camera (Zyla5.5, Andor) and a 100 $\times$ objective with an NA of 0.73 and a working distance of $\SI{4.7}{mm}$ after passing through a long-wavelength optical pass filter with a cutoff wavelength of $\SI{650}{nm}$.
A double-balanced mixer (IQ-1545, Marki) upconverts the baseband I and Q pulses from an arbitrary waveform generator (33622A, Keysight) by mixing with a microwave from a local oscillator (SMC100A, Rhodes-Schwarz).
The upconverted signals are amplified with an amplifier (ZHL-16W-43+, Mini-Circuits) and fed to a microwave planar ring antenna \cite{sasaki2016broadband} placed above a diamond chip. The antenna applys a spatially uniform microwave field in the field of view of the microscope image.
The microwave $\pi/2$ pulse length was $\SI{10}{ns}$.
A pulse sequencer (Pulse Blaster ESR Pro, Spincore) drove the pulsed laser diode, the arbitrary wave generator (33622A, Keysight), the microwave switch, and the scientific CMOS camera.
RF from an RF generator (33120A, Keysight) was applied to a $\SI{10}{\mu m}$ wide Au/Cr wire on a Si chip \cite{mariani2020system, SpringerHybrid}.

\section{Results}

\begin{figure*}[htbp]
\begin{center}
  \includegraphics[width=130mm]{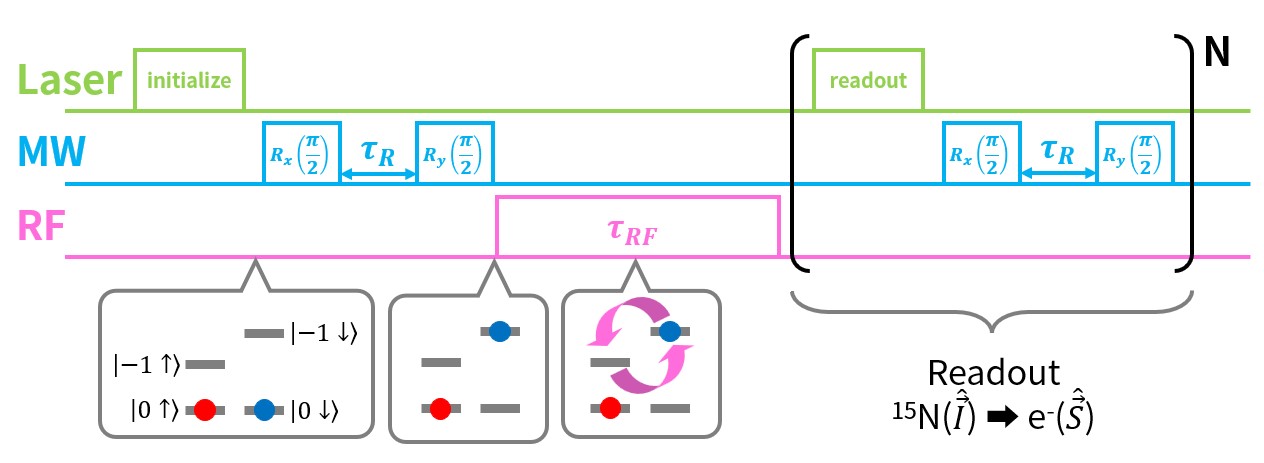}
  \caption{Schematics of a pulse sequence for $^{15}$N nuclear spin Rabi oscillation and Larmor precession. The Bloch sphere on the top of the figure is represented in the rotating coordinate system at $f_M = (f_{\uparrow}+f_{\downarrow})/2$. The NV center is initialized to $\ket{0}$ by a green laser pulse. The electron spin is rotated by $\pi/2$ around the $x$-axis by irradiating with a microwave pulse at $f_M$. The electron spins precess around the $z$-axis by $\pi/2$ at $f_{\uparrow}-f_M$ for $\ket{\uparrow}$ and $f_{\downarrow}-f_M$ for $\ket{\downarrow}$ . The second microwave pulse rotates the electron spin by $\pi/2$ around the $y$-axis. Then Rabi oscillation of $^{15}$N nuclear spins is induced by RF at a resonance frequency $f_R$. Finally, the nuclear spin state is transferred to the electron spins without destroying the $^{15}$N nuclear spin state. The electron spin state is measured by excitation by a green laser pulse that follows.}
  \label{fig:3}
\end{center}
\end{figure*}

The application of the magnetic field leads to a splitting of the degenerate energy levels into the states $\ket{m_S, m_I}$, where $m_S(=0, \pm 1)$ and $m_I(=\uparrow,\downarrow)$ are electron spin and $^{15}$N nuclear spin magnetic quantum numbers, respectively.
Two resonant frequencies $f_{\uparrow}$ for $\ket{m_S=0,m_I=\uparrow} \leftrightarrow \ket{-1,\uparrow}$ and $f_{\downarrow}$ for $\ket{0,\downarrow} \leftrightarrow \ket{-1,\downarrow}$ are determined by pulsed optically detected magnetic resonance (ODMR).

We measure Rabi oscillations of $^{15}$N nuclear spins by varying the angle $\theta$ between the magnetic field and the axis of the NV.
Figure \ref{fig:3} shows a schematics of a pulse sequence
for observation of $^{15}$N nuclear spin Rabi oscillation and Larmor precession. 
We use non-selective pulses for the separation of nuclear spins by state \cite{Azuma22}. 

We consider a rotating coordinate system that rotates at a frequency $f_M = (f_{\uparrow}+f_{\downarrow})/2$.
First, the electron spins are initialized to the state $m_S=0$ by a green laser pulse.
Then, the electron spins are rotated by $\pi/2$ around the $x$-axis by irradiating MW at the frequency $f_M$.
The application of off-resonance pulses induces precession of the electron spins around the $z$-axis. 
Depending on the direction of the electron spin, the electrons in the states $\ket{\uparrow}$ and $\ket{\downarrow}$ precess around the $z$-axis at $f_{\uparrow}-f_M$ or $f_{\downarrow}-f_M$.
Note here that the directions of the rotations are opposite because the sign of the frequencies $f_{\uparrow}-f_M$ and $f_{\downarrow}-f_M$ are different.
Then, a MW pulse at frequency $f_M$ is applied to rotate the electron spins by $\pi/2$ around the $y$-axis.
The above procedure of MW pulse irradiation and precession of the electron spins is called a Ramsey process.
Next, Rabi oscillations of $^{15}$N nuclear spins are induced by an RF pulse at the resonant frequency $f_R$ of $\ket{-1,\uparrow} \leftrightarrow \ket{-1,\downarrow}$.
Finally, the spin state of the $^{15}$N nuclear spins is read after initializing the electron spin by a green laser pulse. 
The Ramsey process is applied again, which transfers the $^{15}$N spin state to the electron spin state. 
Then the electron spin state is read out by the PL by a green laser pulse excitation. The PL intensity reflects the $^{15}$N spin state projected to the $z$-axis. 
In our measurement, the number of readouts $N$ was set to 4.

At a magnetic field angle $\theta = 0^{\circ}$, the rotation axis of the nuclear spins is parallel to the $z$-axis, and hence a Larmor precession of the $^{15}$N nuclear spins is not observed.
In the case of $\theta \neq 0^{\circ}$, 
the oscillation of the $^{15}$N spin state projected to the $z$-axis is observed
because a Larmor precession of the $^{15}$N nuclear spins occurs around the static magnetic field axis away from the $z$-axis.

\begin{figure*}[htbp]
\begin{center}
  \includegraphics[width=130mm]{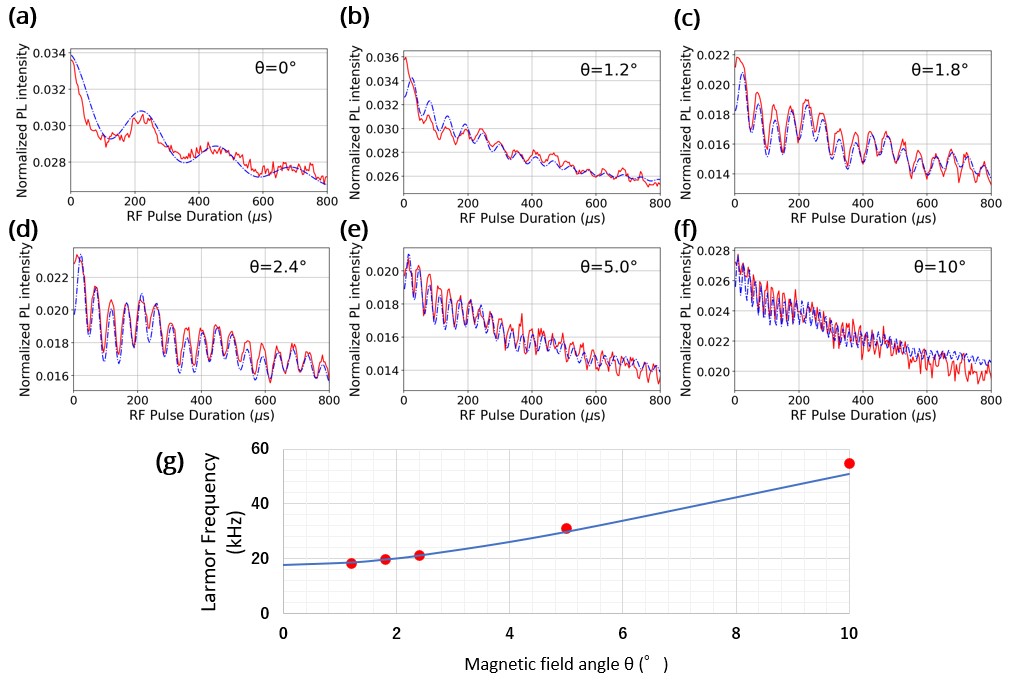}
  \caption{(a)-(f) Measured results by varying the angle between the axis of the NV center and the external static magnetic field ($\theta$) between 0-10$^{\circ}$, showing $^{15}$N nuclear spin Rabi oscillation and Larmor precession. The solid red curves are experimentally obtained results. The blue dashed curves are the best fitted curves to Eq. (\ref{eq:calc}). 
   Only Rabi oscillations of the $^{15}$N nuclear spins are observed at $\theta=0^{\circ}$, while both the Rabi oscillation and the Larmor precession are observed at $\theta \neq 0^{\circ}$. (g) The Larmor precession frequency of the $^{15}$N nuclear spin as a function of the angle $\theta$. The red points are experimentally obtained values and the blue curve is the theoretical value calculated from Eq. (\ref{eq:ms}) without any fitting parameters.}
  \label{fig:4}
\end{center}
\end{figure*}

The observed dynamics of  $^{15}$N nuclear spins are shown in Fig. 4. 
The frequency $f_M$ was adjusted at each angle $\theta$ by observing pulsed-ODMR spectra.
The frequency of the RF pulses were determined from ODMR spectra as a function of RF frequency to be $f_R =\SI{3.012}{MHz}$.
The Larmor frequency for the magnetic quantum number $m_S = -1$ is estimated from Eq. (\ref{eq:ms}) to be $\SI{3.012}{MHz}$ with small variations between $\theta=0^{\circ}$ and $10^{\circ}$, in agreement with the experimentally obtained value.
The data points in Fig. 4 were measured at the RF pulse duration integer multiple of a period of  $1/3.012 {\mu s}$ in order to eliminate the effect of the rapid Larmor oscillation at $m_S = -1$.

The best fitted curves to
\begin{align}
  f(t) = A e^{-Bt} \qty( 1 - C \cos(\omega_R t) - D \cos(\omega_L t) ) + E.
  \label{eq:calc}
\end{align}
are also shown in Figs. 4(a)- 4(f).
The Rabi frequency $\omega_R$ and the Larmor frequency $\omega_L$ are obtained from the curve fittings.

The Rabi frequency $\omega_R$ obtained from the fittings is $2\pi \times \SI{4.30}{kHz}$ for all the traces at $0^{\circ} \le \theta \le 10^{\circ}$ in Figs. 4(a)-4(f). 
This is reasonable because the incident RF amplitude was kept constant.
On the other hand, a small change in $\theta$ is found to change the Larmor frequencies $\omega_L$ significantly as shown in Fig. \ref{fig:4}(g).
The experimentally obtained Larmor frequencies and a theoretical curve (Eq. (\ref{eq:ms})) agree well without any fitted parameters.

\section{Discussion}

We have shown that the Larmor precession of the nuclear spins changes depending on the electron spin state due to the hyperfine interaction. A small change in the angle ($\delta\theta = 10^{\circ}$) between the axis of the NV center and the external static magnetic field leads to a significant change in the Larmor frequency of the nuclear spins from 18 to 55 kHz due to the anisotropy in the hyperfine interaction.
The change in the Larmor frequency of the nuclear spin may be used to measure the magnitude of the lateral static magnetic field at high sensitivity.

The electron-spin-nuclear-spin hybrid system has been experimentally demonstrated to allow highly sensitive ac magnetic field sensing~\cite{rosskopf2017quantum, Pfender17ncomm}.
The coherence time of the nuclear spins is typically $10^3$ times longer than that of the electron spins.
High sensitivity is achieved by transferring the electronic spin state to the nuclear spin state and holding it for a long time limited by $T_{1, n}\approx 52$ ms~\cite{rosskopf2017quantum}.

We propose to utilize the change of the Larmor frequency of the nuclear spins due to the lateral magnetic field to detect static magnetic fields at high sensitivity.
From Eq. (\ref{eq:ms}), the Larmor frequency of  $^{15}$N nuclear spins at $m_S = 0$ is given by 

\begin{align}
  \omega_{m_S=0} =  - \gamma_n B_z \sqrt{ 1 + {\rm tan}^2 \theta \Bigl(1 + 2 \frac{\gamma_e A_{\perp}}{\gamma_n D}\Bigr)
  }.
\end{align}

The accumulated phase at an optimum precession period $T_{2,n}^{*}/2$ is given by
\begin{align}
\omega_{m_S=0} T_{2,n}^{*}/2
\approx  - \gamma_n B_{z} \sqrt{ 1 + 52.4 {\rm tan}^2\theta}T_{2,n}^{*}/2,
\label{eq:accphase}
\end{align}
where the coherence time of $^{15}$N ($T_{2,n}^{*}$) was estimated to be 9 ms.~\cite{Aslam17}
This surpasses the accumulated phase of the NV electron spin by a Ramsey measurement as given by  $\omega_{m_e} T_{2}^{*}/2 = \gamma_e B_{z} T_{2}^{*}/2$
where $T_{2}^{*}$ is typically 1 $\mu$s.
The superiority of the method to utilize $^{15}$N nuclear spin rises sharply as $\theta$ increases because of the large coefficient 52.4 in Eq. (\ref{eq:accphase}),
which leads to further lowering of the minimum detectable magnetic fields.
Moreover, this method has the advantage of being able to read out the accumulated phase of the nuclear spins repeatedly by transferring it to the electron spins.
The coherence of $^{15}$N nuclear spin is disturbed by the spin flip of the NV electron spin, and hence, is limited by the population decay time $T_{1}$ of the NV electron spin. $T_{2,n}^{*}$ can be lengthened by decreasing the lattice temperature or by optical pumping into the $m_s = 0$ electron spin state by 594 nm laser illumination. \cite{Pfender17Nonvolatile}
Whereas the interrogation time and the contrast of read-out have to be taken into consideration practically, the above comparison indicates that the nuclear spin-based measurement method is promising.\\

\section{Concluding Remarks}

Polarization and initialization of $^{15}$N nuclear spins in a short time have been demonstrated by utilizing a method by separating the nitrogen nuclear spins using nonselective microwave pulses. 
The Larmor frequency of the $^{15}$N nuclear spins has been measured by changing the angle $\theta$ between the axis of the NV center and the external magnetic field.
The  Larmor frequency changes significantly with an increase in the angle $\theta$ due to the hyperfine interaction in accordance with a model calculation without any fitting parameters.
We propose a method to lower the minimum detectable static magnetic fields by the Larmor precession of the $^{15}$N nuclear spins.
Our results may contribute to a wide range of applications, such  as magnetic field sensing and quantum information processing using nuclear spin. An accurate understanding of the behavior of the nuclear spin in the presence of the hyperfine interactions contributes to enhancing the sensitivity of quantum sensing and devising novel protocols using NV centers in diamonds.

\begin{acknowledgments}
This work was partly supported by a Grant-in-Aid for Scientific Research (Nos. 21H01009 and 22K18710) from Japan Society for the Promotion of Science. 
\end{acknowledgments}

\bibliography{new}

\begin{thebibliography}{25}%
\makeatletter
\providecommand \@ifxundefined [1]{%
 \@ifx{#1\undefined}
}%
\providecommand \@ifnum [1]{%
 \ifnum #1\expandafter \@firstoftwo
 \else \expandafter \@secondoftwo
 \fi
}%
\providecommand \@ifx [1]{%
 \ifx #1\expandafter \@firstoftwo
 \else \expandafter \@secondoftwo
 \fi
}%
\providecommand \natexlab [1]{#1}%
\providecommand \enquote  [1]{``#1''}%
\providecommand \bibnamefont  [1]{#1}%
\providecommand \bibfnamefont [1]{#1}%
\providecommand \citenamefont [1]{#1}%
\providecommand \href@noop [0]{\@secondoftwo}%
\providecommand \href [0]{\begingroup \@sanitize@url \@href}%
\providecommand \@href[1]{\@@startlink{#1}\@@href}%
\providecommand \@@href[1]{\endgroup#1\@@endlink}%
\providecommand \@sanitize@url [0]{\catcode `\\12\catcode `\$12\catcode
  `\&12\catcode `\#12\catcode `\^12\catcode `\_12\catcode `\%12\relax}%
\providecommand \@@startlink[1]{}%
\providecommand \@@endlink[0]{}%
\providecommand \url  [0]{\begingroup\@sanitize@url \@url }%
\providecommand \@url [1]{\endgroup\@href {#1}{\urlprefix }}%
\providecommand \urlprefix  [0]{URL }%
\providecommand \Eprint [0]{\href }%
\providecommand \doibase [0]{https://doi.org/}%
\providecommand \selectlanguage [0]{\@gobble}%
\providecommand \bibinfo  [0]{\@secondoftwo}%
\providecommand \bibfield  [0]{\@secondoftwo}%
\providecommand \translation [1]{[#1]}%
\providecommand \BibitemOpen [0]{}%
\providecommand \bibitemStop [0]{}%
\providecommand \bibitemNoStop [0]{.\EOS\space}%
\providecommand \EOS [0]{\spacefactor3000\relax}%
\providecommand \BibitemShut  [1]{\csname bibitem#1\endcsname}%
\let\auto@bib@innerbib\@empty
\bibitem [{\citenamefont {Degen}\ \emph {et~al.}(2017)\citenamefont {Degen},
  \citenamefont {Reinhard},\ and\ \citenamefont
  {Cappellaro}}]{degen2017quantum}%
  \BibitemOpen
  \bibfield  {author} {\bibinfo {author} {\bibfnamefont {C.~L.}\ \bibnamefont
  {Degen}}, \bibinfo {author} {\bibfnamefont {F.}~\bibnamefont {Reinhard}},\
  and\ \bibinfo {author} {\bibfnamefont {P.}~\bibnamefont {Cappellaro}},\
  }\bibfield  {title} {\bibinfo {title} {Quantum sensing},\ }\href@noop {}
  {\bibfield  {journal} {\bibinfo  {journal} {Rev. Mod. Phys.}\ }\textbf
  {\bibinfo {volume} {89}},\ \bibinfo {pages} {035002} (\bibinfo {year}
  {2017})}\BibitemShut {NoStop}%
\bibitem [{\citenamefont {Doherty}\ \emph {et~al.}(2013)\citenamefont
  {Doherty}, \citenamefont {Manson}, \citenamefont {Delaney}, \citenamefont
  {Jelezko}, \citenamefont {Wrachtrup},\ and\ \citenamefont
  {Hollenberg}}]{Doherty_2013}%
  \BibitemOpen
  \bibfield  {author} {\bibinfo {author} {\bibfnamefont {M.~W.}\ \bibnamefont
  {Doherty}}, \bibinfo {author} {\bibfnamefont {N.~B.}\ \bibnamefont {Manson}},
  \bibinfo {author} {\bibfnamefont {P.}~\bibnamefont {Delaney}}, \bibinfo
  {author} {\bibfnamefont {F.}~\bibnamefont {Jelezko}}, \bibinfo {author}
  {\bibfnamefont {J.}~\bibnamefont {Wrachtrup}},\ and\ \bibinfo {author}
  {\bibfnamefont {L.~C.}\ \bibnamefont {Hollenberg}},\ }\bibfield  {title}
  {\bibinfo {title} {The {n}itrogen-{v}acancy colour centre in diamond},\
  }\href@noop {} {\bibfield  {journal} {\bibinfo  {journal} {Phys. Rep.}\
  }\textbf {\bibinfo {volume} {528}},\ \bibinfo {pages} {1} (\bibinfo {year}
  {2013})}\BibitemShut {NoStop}%
\bibitem [{\citenamefont {Barry}\ \emph {et~al.}(2020)\citenamefont {Barry},
  \citenamefont {Schloss}, \citenamefont {Bauch}, \citenamefont {Turner},
  \citenamefont {Hart}, \citenamefont {Pham},\ and\ \citenamefont
  {Walsworth}}]{Barry20}%
  \BibitemOpen
  \bibfield  {author} {\bibinfo {author} {\bibfnamefont {J.~F.}\ \bibnamefont
  {Barry}}, \bibinfo {author} {\bibfnamefont {J.~M.}\ \bibnamefont {Schloss}},
  \bibinfo {author} {\bibfnamefont {E.}~\bibnamefont {Bauch}}, \bibinfo
  {author} {\bibfnamefont {M.~J.}\ \bibnamefont {Turner}}, \bibinfo {author}
  {\bibfnamefont {C.~A.}\ \bibnamefont {Hart}}, \bibinfo {author}
  {\bibfnamefont {L.~M.}\ \bibnamefont {Pham}},\ and\ \bibinfo {author}
  {\bibfnamefont {R.~L.}\ \bibnamefont {Walsworth}},\ }\bibfield  {title}
  {\bibinfo {title} {Sensitivity optimization for nv-diamond magnetometry},\
  }\href {https://doi.org/10.1103/RevModPhys.92.015004} {\bibfield  {journal}
  {\bibinfo  {journal} {Rev. Mod. Phys.}\ }\textbf {\bibinfo {volume} {92}},\
  \bibinfo {pages} {015004} (\bibinfo {year} {2020})}\BibitemShut {NoStop}%
\bibitem [{\citenamefont {Jiang}\ \emph {et~al.}(2009)\citenamefont {Jiang},
  \citenamefont {Hodges}, \citenamefont {Maze}, \citenamefont {Maurer},
  \citenamefont {Taylor}, \citenamefont {Cory}, \citenamefont {Hemmer},
  \citenamefont {Walsworth}, \citenamefont {Yacoby}, \citenamefont {Zibrov},\
  and\ \citenamefont {Lukin}}]{Jiang09}%
  \BibitemOpen
  \bibfield  {author} {\bibinfo {author} {\bibfnamefont {L.}~\bibnamefont
  {Jiang}}, \bibinfo {author} {\bibfnamefont {J.~S.}\ \bibnamefont {Hodges}},
  \bibinfo {author} {\bibfnamefont {J.~R.}\ \bibnamefont {Maze}}, \bibinfo
  {author} {\bibfnamefont {P.}~\bibnamefont {Maurer}}, \bibinfo {author}
  {\bibfnamefont {J.~M.}\ \bibnamefont {Taylor}}, \bibinfo {author}
  {\bibfnamefont {D.~G.}\ \bibnamefont {Cory}}, \bibinfo {author}
  {\bibfnamefont {P.~R.}\ \bibnamefont {Hemmer}}, \bibinfo {author}
  {\bibfnamefont {R.~L.}\ \bibnamefont {Walsworth}}, \bibinfo {author}
  {\bibfnamefont {A.}~\bibnamefont {Yacoby}}, \bibinfo {author} {\bibfnamefont
  {A.~S.}\ \bibnamefont {Zibrov}},\ and\ \bibinfo {author} {\bibfnamefont
  {M.~D.}\ \bibnamefont {Lukin}},\ }\bibfield  {title} {\bibinfo {title}
  {Repetitive readout of a single electronic spin via quantum logic with
  nuclear spin ancillae},\ }\href {https://doi.org/10.1126/science.1176496}
  {\bibfield  {journal} {\bibinfo  {journal} {Science}\ }\textbf {\bibinfo
  {volume} {326}},\ \bibinfo {pages} {267} (\bibinfo {year}
  {2009})}\BibitemShut {NoStop}%
\bibitem [{\citenamefont {Zaiser}\ \emph {et~al.}(2016)\citenamefont {Zaiser},
  \citenamefont {Rendler}, \citenamefont {Jakobi}, \citenamefont {Wolf},
  \citenamefont {Lee}, \citenamefont {Wagner}, \citenamefont {Bergholm},
  \citenamefont {Schulte-Herbrüggen}, \citenamefont {Neumann},\ and\
  \citenamefont {Wrachtrup}}]{Zaiser16}%
  \BibitemOpen
  \bibfield  {author} {\bibinfo {author} {\bibfnamefont {S.}~\bibnamefont
  {Zaiser}}, \bibinfo {author} {\bibfnamefont {T.}~\bibnamefont {Rendler}},
  \bibinfo {author} {\bibfnamefont {I.}~\bibnamefont {Jakobi}}, \bibinfo
  {author} {\bibfnamefont {T.}~\bibnamefont {Wolf}}, \bibinfo {author}
  {\bibfnamefont {S.-Y.}\ \bibnamefont {Lee}}, \bibinfo {author} {\bibfnamefont
  {S.}~\bibnamefont {Wagner}}, \bibinfo {author} {\bibfnamefont
  {V.}~\bibnamefont {Bergholm}}, \bibinfo {author} {\bibfnamefont
  {T.}~\bibnamefont {Schulte-Herbrüggen}}, \bibinfo {author} {\bibfnamefont
  {P.}~\bibnamefont {Neumann}},\ and\ \bibinfo {author} {\bibfnamefont
  {J.}~\bibnamefont {Wrachtrup}},\ }\bibfield  {title} {\bibinfo {title} {A
  quantum spectrum analyzer enhanced by a nuclear spin memory},\ }\href@noop {}
  {\bibfield  {journal} {\bibinfo  {journal} {Nat. Comm.}\ }\textbf {\bibinfo
  {volume} {7}},\ \bibinfo {pages} {12279} (\bibinfo {year}
  {2016})}\BibitemShut {NoStop}%
\bibitem [{\citenamefont {Matsuzaki}\ \emph {et~al.}(2016)\citenamefont
  {Matsuzaki}, \citenamefont {Shimo-Oka}, \citenamefont {Tanaka}, \citenamefont
  {Tokura}, \citenamefont {Semba},\ and\ \citenamefont
  {Mizuochi}}]{matsuzaki2016hybrid}%
  \BibitemOpen
  \bibfield  {author} {\bibinfo {author} {\bibfnamefont {Y.}~\bibnamefont
  {Matsuzaki}}, \bibinfo {author} {\bibfnamefont {T.}~\bibnamefont
  {Shimo-Oka}}, \bibinfo {author} {\bibfnamefont {H.}~\bibnamefont {Tanaka}},
  \bibinfo {author} {\bibfnamefont {Y.}~\bibnamefont {Tokura}}, \bibinfo
  {author} {\bibfnamefont {K.}~\bibnamefont {Semba}},\ and\ \bibinfo {author}
  {\bibfnamefont {N.}~\bibnamefont {Mizuochi}},\ }\bibfield  {title} {\bibinfo
  {title} {Hybrid quantum magnetic-field sensor with an electron spin and a
  nuclear spin in diamond},\ }\href@noop {} {\bibfield  {journal} {\bibinfo
  {journal} {Phys. Rev. A}\ }\textbf {\bibinfo {volume} {94}},\ \bibinfo
  {pages} {052330} (\bibinfo {year} {2016})}\BibitemShut {NoStop}%
\bibitem [{\citenamefont {Rosskopf}\ \emph {et~al.}(2017)\citenamefont
  {Rosskopf}, \citenamefont {Zopes}, \citenamefont {Boss},\ and\ \citenamefont
  {Degen}}]{rosskopf2017quantum}%
  \BibitemOpen
  \bibfield  {author} {\bibinfo {author} {\bibfnamefont {T.}~\bibnamefont
  {Rosskopf}}, \bibinfo {author} {\bibfnamefont {J.}~\bibnamefont {Zopes}},
  \bibinfo {author} {\bibfnamefont {J.~M.}\ \bibnamefont {Boss}},\ and\
  \bibinfo {author} {\bibfnamefont {C.~L.}\ \bibnamefont {Degen}},\ }\bibfield
  {title} {\bibinfo {title} {A quantum spectrum analyzer enhanced by a nuclear
  spin memory},\ }\href@noop {} {\bibfield  {journal} {\bibinfo  {journal} {npj
  Quantum Information}\ }\textbf {\bibinfo {volume} {3}},\ \bibinfo {pages} {1}
  (\bibinfo {year} {2017})}\BibitemShut {NoStop}%
\bibitem [{\citenamefont {Pfender}\ \emph
  {et~al.}(2017{\natexlab{a}})\citenamefont {Pfender}, \citenamefont {Aslam},
  \citenamefont {Sumiya}, \citenamefont {Onoda}, \citenamefont {Neumann},
  \citenamefont {Isoya}, \citenamefont {Meriles},\ and\ \citenamefont
  {Wrachtrup}}]{Pfender17ncomm}%
  \BibitemOpen
  \bibfield  {author} {\bibinfo {author} {\bibfnamefont {M.}~\bibnamefont
  {Pfender}}, \bibinfo {author} {\bibfnamefont {N.}~\bibnamefont {Aslam}},
  \bibinfo {author} {\bibfnamefont {H.}~\bibnamefont {Sumiya}}, \bibinfo
  {author} {\bibfnamefont {S.}~\bibnamefont {Onoda}}, \bibinfo {author}
  {\bibfnamefont {P.}~\bibnamefont {Neumann}}, \bibinfo {author} {\bibfnamefont
  {J.}~\bibnamefont {Isoya}}, \bibinfo {author} {\bibfnamefont {C.~A.}\
  \bibnamefont {Meriles}},\ and\ \bibinfo {author} {\bibfnamefont
  {J.}~\bibnamefont {Wrachtrup}},\ }\bibfield  {title} {\bibinfo {title}
  {Protecting a diamond quantum memory by charge state control},\ }\href@noop
  {} {\bibfield  {journal} {\bibinfo  {journal} {Nat. Comm.}\ }\textbf
  {\bibinfo {volume} {7}},\ \bibinfo {pages} {12279} (\bibinfo {year}
  {2017}{\natexlab{a}})}\BibitemShut {NoStop}%
\bibitem [{\citenamefont {Felton}\ \emph {et~al.}(2009)\citenamefont {Felton},
  \citenamefont {Edmonds}, \citenamefont {Newton}, \citenamefont {Martineau},
  \citenamefont {Fisher}, \citenamefont {Twitchen},\ and\ \citenamefont
  {Baker}}]{PhysRevB.79.075203}%
  \BibitemOpen
  \bibfield  {author} {\bibinfo {author} {\bibfnamefont {S.}~\bibnamefont
  {Felton}}, \bibinfo {author} {\bibfnamefont {A.~M.}\ \bibnamefont {Edmonds}},
  \bibinfo {author} {\bibfnamefont {M.~E.}\ \bibnamefont {Newton}}, \bibinfo
  {author} {\bibfnamefont {P.~M.}\ \bibnamefont {Martineau}}, \bibinfo {author}
  {\bibfnamefont {D.}~\bibnamefont {Fisher}}, \bibinfo {author} {\bibfnamefont
  {D.~J.}\ \bibnamefont {Twitchen}},\ and\ \bibinfo {author} {\bibfnamefont
  {J.~M.}\ \bibnamefont {Baker}},\ }\bibfield  {title} {\bibinfo {title}
  {Hyperfine interaction in the ground state of the negatively charged
  {n}itrogen {v}acancy center in diamond},\ }\href@noop {} {\bibfield
  {journal} {\bibinfo  {journal} {Phys. Rev. B}\ }\textbf {\bibinfo {volume}
  {79}},\ \bibinfo {pages} {075203} (\bibinfo {year} {2009})}\BibitemShut
  {NoStop}%
\bibitem [{\citenamefont {He}\ \emph {et~al.}(1993)\citenamefont {He},
  \citenamefont {Manson},\ and\ \citenamefont {Fisk}}]{PhysRevB.47.8809}%
  \BibitemOpen
  \bibfield  {author} {\bibinfo {author} {\bibfnamefont {X.-F.}\ \bibnamefont
  {He}}, \bibinfo {author} {\bibfnamefont {N.~B.}\ \bibnamefont {Manson}},\
  and\ \bibinfo {author} {\bibfnamefont {P.~T.~H.}\ \bibnamefont {Fisk}},\
  }\bibfield  {title} {\bibinfo {title} {Paramagnetic resonance of
  photoexcited{N-V} defects in diamond. i. level anticrossing in the $^{3}$a
  ground state},\ }\href {https://doi.org/10.1103/PhysRevB.47.8809} {\bibfield
  {journal} {\bibinfo  {journal} {Phys. Rev. B}\ }\textbf {\bibinfo {volume}
  {47}},\ \bibinfo {pages} {8809} (\bibinfo {year} {1993})}\BibitemShut
  {NoStop}%
\bibitem [{\citenamefont {Jacques}\ \emph {et~al.}(2009)\citenamefont
  {Jacques}, \citenamefont {Neumann}, \citenamefont {Beck}, \citenamefont
  {Markham}, \citenamefont {Twitchen}, \citenamefont {Meijer}, \citenamefont
  {Kaiser}, \citenamefont {Balasubramanian}, \citenamefont {Jelezko},\ and\
  \citenamefont {Wrachtrup}}]{PhysRevLett.102.057403}%
  \BibitemOpen
  \bibfield  {author} {\bibinfo {author} {\bibfnamefont {V.}~\bibnamefont
  {Jacques}}, \bibinfo {author} {\bibfnamefont {P.}~\bibnamefont {Neumann}},
  \bibinfo {author} {\bibfnamefont {J.}~\bibnamefont {Beck}}, \bibinfo {author}
  {\bibfnamefont {M.}~\bibnamefont {Markham}}, \bibinfo {author} {\bibfnamefont
  {D.}~\bibnamefont {Twitchen}}, \bibinfo {author} {\bibfnamefont
  {J.}~\bibnamefont {Meijer}}, \bibinfo {author} {\bibfnamefont
  {F.}~\bibnamefont {Kaiser}}, \bibinfo {author} {\bibfnamefont
  {G.}~\bibnamefont {Balasubramanian}}, \bibinfo {author} {\bibfnamefont
  {F.}~\bibnamefont {Jelezko}},\ and\ \bibinfo {author} {\bibfnamefont
  {J.}~\bibnamefont {Wrachtrup}},\ }\bibfield  {title} {\bibinfo {title}
  {Dynamic polarization of single nuclear spins by optical pumping of
  {N}itrogen-{V}acancy {C}olor {C}enters in diamond at room temperature},\
  }\href {https://doi.org/10.1103/PhysRevLett.102.057403} {\bibfield  {journal}
  {\bibinfo  {journal} {Phys. Rev. Lett.}\ }\textbf {\bibinfo {volume} {102}},\
  \bibinfo {pages} {057403} (\bibinfo {year} {2009})}\BibitemShut {NoStop}%
\bibitem [{\citenamefont {Busaite}\ \emph {et~al.}(2020)\citenamefont
  {Busaite}, \citenamefont {Lazda}, \citenamefont {Berzins}, \citenamefont
  {Auzinsh}, \citenamefont {Ferber},\ and\ \citenamefont
  {Gahbauer}}]{PhysRevB.102.224101}%
  \BibitemOpen
  \bibfield  {author} {\bibinfo {author} {\bibfnamefont {L.}~\bibnamefont
  {Busaite}}, \bibinfo {author} {\bibfnamefont {R.}~\bibnamefont {Lazda}},
  \bibinfo {author} {\bibfnamefont {A.}~\bibnamefont {Berzins}}, \bibinfo
  {author} {\bibfnamefont {M.}~\bibnamefont {Auzinsh}}, \bibinfo {author}
  {\bibfnamefont {R.}~\bibnamefont {Ferber}},\ and\ \bibinfo {author}
  {\bibfnamefont {F.}~\bibnamefont {Gahbauer}},\ }\bibfield  {title} {\bibinfo
  {title} {Dynamic 14{N} nuclear spin polarization in {n}itrogen-{v}acancy
  centers in diamond},\ }\href {https://doi.org/10.1103/PhysRevB.102.224101}
  {\bibfield  {journal} {\bibinfo  {journal} {Phys. Rev. B}\ }\textbf {\bibinfo
  {volume} {102}},\ \bibinfo {pages} {224101} (\bibinfo {year}
  {2020})}\BibitemShut {NoStop}%
\bibitem [{\citenamefont {Morishita}\ \emph {et~al.}(2020)\citenamefont
  {Morishita}, \citenamefont {Kobayashi}, \citenamefont {Fujiwara},
  \citenamefont {Kato}, \citenamefont {Makino}, \citenamefont {Yamasaki},\ and\
  \citenamefont {Mizuochi}}]{Morishita20}%
  \BibitemOpen
  \bibfield  {author} {\bibinfo {author} {\bibfnamefont {H.}~\bibnamefont
  {Morishita}}, \bibinfo {author} {\bibfnamefont {S.}~\bibnamefont
  {Kobayashi}}, \bibinfo {author} {\bibfnamefont {M.}~\bibnamefont {Fujiwara}},
  \bibinfo {author} {\bibfnamefont {H.}~\bibnamefont {Kato}}, \bibinfo {author}
  {\bibfnamefont {T.}~\bibnamefont {Makino}}, \bibinfo {author} {\bibfnamefont
  {S.}~\bibnamefont {Yamasaki}},\ and\ \bibinfo {author} {\bibfnamefont
  {N.}~\bibnamefont {Mizuochi}},\ }\bibfield  {title} {\bibinfo {title} {Room
  temperature electrically detected nuclear spin coherence of nv centres in
  diamond},\ }\href@noop {} {\bibfield  {journal} {\bibinfo  {journal} {Sci.
  Rep.}\ }\textbf {\bibinfo {volume} {10}},\ \bibinfo {pages} {792} (\bibinfo
  {year} {2020})}\BibitemShut {NoStop}%
\bibitem [{\citenamefont {Gulka}\ \emph {et~al.}(2021)\citenamefont {Gulka},
  \citenamefont {Wirtitsch}, \citenamefont {Ivády}, \citenamefont {Vodnik},
  \citenamefont {Hruby}, \citenamefont {Magchiels}, \citenamefont {Bourgeois},
  \citenamefont {Gali}, \citenamefont {Trupke},\ and\ \citenamefont
  {Nesladek}}]{Gulka21}%
  \BibitemOpen
  \bibfield  {author} {\bibinfo {author} {\bibfnamefont {M.}~\bibnamefont
  {Gulka}}, \bibinfo {author} {\bibfnamefont {D.}~\bibnamefont {Wirtitsch}},
  \bibinfo {author} {\bibfnamefont {V.}~\bibnamefont {Ivády}}, \bibinfo
  {author} {\bibfnamefont {J.}~\bibnamefont {Vodnik}}, \bibinfo {author}
  {\bibfnamefont {J.}~\bibnamefont {Hruby}}, \bibinfo {author} {\bibfnamefont
  {G.}~\bibnamefont {Magchiels}}, \bibinfo {author} {\bibfnamefont
  {E.}~\bibnamefont {Bourgeois}}, \bibinfo {author} {\bibfnamefont
  {A.}~\bibnamefont {Gali}}, \bibinfo {author} {\bibfnamefont {M.}~\bibnamefont
  {Trupke}},\ and\ \bibinfo {author} {\bibfnamefont {M.}~\bibnamefont
  {Nesladek}},\ }\bibfield  {title} {\bibinfo {title} {Room-temperature control
  and electrical readout of individual nitrogen-vacancy nuclear spins},\
  }\href@noop {} {\bibfield  {journal} {\bibinfo  {journal} {Nat. Commun.}\
  }\textbf {\bibinfo {volume} {12}},\ \bibinfo {pages} {4421} (\bibinfo {year}
  {2021})}\BibitemShut {NoStop}%
\bibitem [{\citenamefont {Azuma}\ \emph {et~al.}(2022)\citenamefont {Azuma},
  \citenamefont {Watanabe}, \citenamefont {Kashiwaya},\ and\ \citenamefont
  {Nomura}}]{Azuma22}%
  \BibitemOpen
  \bibfield  {author} {\bibinfo {author} {\bibfnamefont {Y.}~\bibnamefont
  {Azuma}}, \bibinfo {author} {\bibfnamefont {H.}~\bibnamefont {Watanabe}},
  \bibinfo {author} {\bibfnamefont {S.}~\bibnamefont {Kashiwaya}},\ and\
  \bibinfo {author} {\bibfnamefont {S.}~\bibnamefont {Nomura}},\ }\bibfield
  {title} {\bibinfo {title} {Rapid control of $\mathrm{^{15}N}$ nuclear spin
  within diamond nv centers},\ }\href@noop {} {\bibfield  {journal} {\bibinfo
  {journal} {Extended Abstracts of the 2022 International Conference on Solid
  State Devices and Materials}\ ,\ \bibinfo {pages} {195}} (\bibinfo {year}
  {2022})}\BibitemShut {NoStop}%
\bibitem [{\citenamefont {Ofori-Okai}\ \emph {et~al.}(2012)\citenamefont
  {Ofori-Okai}, \citenamefont {Pezzagna}, \citenamefont {Chang}, \citenamefont
  {Loretz}, \citenamefont {Schirhagl}, \citenamefont {Tao}, \citenamefont
  {Moores}, \citenamefont {Groot-Berning}, \citenamefont {Meijer},\ and\
  \citenamefont {Degen}}]{ofori2012spin}%
  \BibitemOpen
  \bibfield  {author} {\bibinfo {author} {\bibfnamefont {B.}~\bibnamefont
  {Ofori-Okai}}, \bibinfo {author} {\bibfnamefont {S.}~\bibnamefont
  {Pezzagna}}, \bibinfo {author} {\bibfnamefont {K.}~\bibnamefont {Chang}},
  \bibinfo {author} {\bibfnamefont {M.}~\bibnamefont {Loretz}}, \bibinfo
  {author} {\bibfnamefont {R.}~\bibnamefont {Schirhagl}}, \bibinfo {author}
  {\bibfnamefont {Y.}~\bibnamefont {Tao}}, \bibinfo {author} {\bibfnamefont
  {B.}~\bibnamefont {Moores}}, \bibinfo {author} {\bibfnamefont
  {K.}~\bibnamefont {Groot-Berning}}, \bibinfo {author} {\bibfnamefont
  {J.}~\bibnamefont {Meijer}},\ and\ \bibinfo {author} {\bibfnamefont
  {C.}~\bibnamefont {Degen}},\ }\bibfield  {title} {\bibinfo {title} {Spin
  properties of very shallow nitrogen vacancy defects in diamond},\ }\href@noop
  {} {\bibfield  {journal} {\bibinfo  {journal} {Phys. Rev. B}\ }\textbf
  {\bibinfo {volume} {86}},\ \bibinfo {pages} {081406} (\bibinfo {year}
  {2012})}\BibitemShut {NoStop}%
\bibitem [{\citenamefont {Mariani}\ \emph {et~al.}(2020)\citenamefont
  {Mariani}, \citenamefont {Nomoto}, \citenamefont {Kashiwaya},\ and\
  \citenamefont {Nomura}}]{mariani2020system}%
  \BibitemOpen
  \bibfield  {author} {\bibinfo {author} {\bibfnamefont {G.}~\bibnamefont
  {Mariani}}, \bibinfo {author} {\bibfnamefont {S.}~\bibnamefont {Nomoto}},
  \bibinfo {author} {\bibfnamefont {S.}~\bibnamefont {Kashiwaya}},\ and\
  \bibinfo {author} {\bibfnamefont {S.}~\bibnamefont {Nomura}},\ }\bibfield
  {title} {\bibinfo {title} {System for the remote control and imaging of mw
  fields for spin manipulation in nv centers in diamond},\ }\href@noop {}
  {\bibfield  {journal} {\bibinfo  {journal} {Sci. Rep.}\ }\textbf {\bibinfo
  {volume} {10}},\ \bibinfo {pages} {1} (\bibinfo {year} {2020})}\BibitemShut
  {NoStop}%
\bibitem [{\citenamefont {Maze}\ \emph {et~al.}(2008)\citenamefont {Maze},
  \citenamefont {Taylor},\ and\ \citenamefont {Lukin}}]{Maze08ElectronSpin}%
  \BibitemOpen
  \bibfield  {author} {\bibinfo {author} {\bibfnamefont {J.~R.}\ \bibnamefont
  {Maze}}, \bibinfo {author} {\bibfnamefont {J.~M.}\ \bibnamefont {Taylor}},\
  and\ \bibinfo {author} {\bibfnamefont {M.~D.}\ \bibnamefont {Lukin}},\
  }\bibfield  {title} {\bibinfo {title} {Electron spin decoherence of single
  nitrogen-vacancy defects in diamond},\ }\href@noop {} {\bibfield  {journal}
  {\bibinfo  {journal} {Phys. Rev. B}\ }\textbf {\bibinfo {volume} {78}},\
  \bibinfo {pages} {094303} (\bibinfo {year} {2008})}\BibitemShut {NoStop}%
\bibitem [{\citenamefont {Childress}\ \emph {et~al.}(2006)\citenamefont
  {Childress}, \citenamefont {Gurudev~Dutt}, \citenamefont {Taylor},
  \citenamefont {Zibrov}, \citenamefont {Jelezko}, \citenamefont {Wrachtrup},
  \citenamefont {Hemmer},\ and\ \citenamefont {Lukin}}]{childress2006coherent}%
  \BibitemOpen
  \bibfield  {author} {\bibinfo {author} {\bibfnamefont {L.}~\bibnamefont
  {Childress}}, \bibinfo {author} {\bibfnamefont {M.}~\bibnamefont
  {Gurudev~Dutt}}, \bibinfo {author} {\bibfnamefont {J.}~\bibnamefont
  {Taylor}}, \bibinfo {author} {\bibfnamefont {A.}~\bibnamefont {Zibrov}},
  \bibinfo {author} {\bibfnamefont {F.}~\bibnamefont {Jelezko}}, \bibinfo
  {author} {\bibfnamefont {J.}~\bibnamefont {Wrachtrup}}, \bibinfo {author}
  {\bibfnamefont {P.}~\bibnamefont {Hemmer}},\ and\ \bibinfo {author}
  {\bibfnamefont {M.}~\bibnamefont {Lukin}},\ }\bibfield  {title} {\bibinfo
  {title} {Coherent dynamics of coupled electron and nuclear spin qubits in
  diamond},\ }\href@noop {} {\bibfield  {journal} {\bibinfo  {journal}
  {Science}\ }\textbf {\bibinfo {volume} {314}},\ \bibinfo {pages} {281}
  (\bibinfo {year} {2006})}\BibitemShut {NoStop}%
\bibitem [{\citenamefont {Oon}\ \emph {et~al.}(2022)\citenamefont {Oon},
  \citenamefont {Tang}, \citenamefont {Hart}, \citenamefont {Olsson},
  \citenamefont {Turner}, \citenamefont {Schloss},\ and\ \citenamefont
  {Walsworth}}]{oon2022ramsey}%
  \BibitemOpen
  \bibfield  {author} {\bibinfo {author} {\bibfnamefont {J.~T.}\ \bibnamefont
  {Oon}}, \bibinfo {author} {\bibfnamefont {J.}~\bibnamefont {Tang}}, \bibinfo
  {author} {\bibfnamefont {C.}~\bibnamefont {Hart}}, \bibinfo {author}
  {\bibfnamefont {K.}~\bibnamefont {Olsson}}, \bibinfo {author} {\bibfnamefont
  {M.}~\bibnamefont {Turner}}, \bibinfo {author} {\bibfnamefont
  {J.}~\bibnamefont {Schloss}},\ and\ \bibinfo {author} {\bibfnamefont
  {R.}~\bibnamefont {Walsworth}},\ }\bibfield  {title} {\bibinfo {title}
  {Ramsey envelope modulation in {NV} diamond magnetometry},\ }\href@noop {}
  {\bibfield  {journal} {\bibinfo  {journal} {Bulletin of the American Physical
  Society}\ } (\bibinfo {year} {2022})}\BibitemShut {NoStop}%
\bibitem [{\citenamefont {Nomura}\ \emph {et~al.}(2021)\citenamefont {Nomura},
  \citenamefont {Kaida}, \citenamefont {Watanabe},\ and\ \citenamefont
  {Kashiwaya}}]{nomura2021near}%
  \BibitemOpen
  \bibfield  {author} {\bibinfo {author} {\bibfnamefont {S.}~\bibnamefont
  {Nomura}}, \bibinfo {author} {\bibfnamefont {K.}~\bibnamefont {Kaida}},
  \bibinfo {author} {\bibfnamefont {H.}~\bibnamefont {Watanabe}},\ and\
  \bibinfo {author} {\bibfnamefont {S.}~\bibnamefont {Kashiwaya}},\ }\bibfield
  {title} {\bibinfo {title} {Near-field radio-frequency imaging by spin-locking
  with a nitrogen-vacancy spin sensor},\ }\href@noop {} {\bibfield  {journal}
  {\bibinfo  {journal} {J. Appl. Phys.}\ }\textbf {\bibinfo {volume} {130}},\
  \bibinfo {pages} {024503} (\bibinfo {year} {2021})}\BibitemShut {NoStop}%
\bibitem [{\citenamefont {Sasaki}\ \emph {et~al.}(2016)\citenamefont {Sasaki},
  \citenamefont {Monnai}, \citenamefont {Saijo}, \citenamefont {Fujita},
  \citenamefont {Watanabe}, \citenamefont {Ishi-Hayase}, \citenamefont {Itoh},\
  and\ \citenamefont {Abe}}]{sasaki2016broadband}%
  \BibitemOpen
  \bibfield  {author} {\bibinfo {author} {\bibfnamefont {K.}~\bibnamefont
  {Sasaki}}, \bibinfo {author} {\bibfnamefont {Y.}~\bibnamefont {Monnai}},
  \bibinfo {author} {\bibfnamefont {S.}~\bibnamefont {Saijo}}, \bibinfo
  {author} {\bibfnamefont {R.}~\bibnamefont {Fujita}}, \bibinfo {author}
  {\bibfnamefont {H.}~\bibnamefont {Watanabe}}, \bibinfo {author}
  {\bibfnamefont {J.}~\bibnamefont {Ishi-Hayase}}, \bibinfo {author}
  {\bibfnamefont {K.~M.}\ \bibnamefont {Itoh}},\ and\ \bibinfo {author}
  {\bibfnamefont {E.}~\bibnamefont {Abe}},\ }\bibfield  {title} {\bibinfo
  {title} {Broadband, large-area microwave antenna for optically detected
  magnetic resonance of nitrogen-vacancy centers in diamond},\ }\href@noop {}
  {\bibfield  {journal} {\bibinfo  {journal} {Rev. Sci. Instrum.}\ }\textbf
  {\bibinfo {volume} {87}},\ \bibinfo {pages} {053904} (\bibinfo {year}
  {2016})}\BibitemShut {NoStop}%
\bibitem [{\citenamefont {Nomura}(2021)}]{SpringerHybrid}%
  \BibitemOpen
  \bibfield  {author} {\bibinfo {author} {\bibfnamefont {S.}~\bibnamefont
  {Nomura}},\ }\bibinfo {title} {Hybrid quantum systems}\ (\bibinfo
  {publisher} {Springer Nature},\ \bibinfo {year} {2021})\ Chap.~\bibinfo
  {chapter} {2}, p.~\bibinfo {pages} {27}\BibitemShut {NoStop}%
\bibitem [{\citenamefont {Aslam}\ \emph {et~al.}(2017)\citenamefont {Aslam},
  \citenamefont {Pfender}, \citenamefont {Neumann}, \citenamefont {Reuter},
  \citenamefont {Zappe}, \citenamefont {de~Oliveira}, \citenamefont
  {Denisenko}, \citenamefont {Sumiya}, \citenamefont {Onoda}, \citenamefont
  {Isoya},\ and\ \citenamefont {Wrachtrup}}]{Aslam17}%
  \BibitemOpen
  \bibfield  {author} {\bibinfo {author} {\bibfnamefont {N.}~\bibnamefont
  {Aslam}}, \bibinfo {author} {\bibfnamefont {M.}~\bibnamefont {Pfender}},
  \bibinfo {author} {\bibfnamefont {P.}~\bibnamefont {Neumann}}, \bibinfo
  {author} {\bibfnamefont {R.}~\bibnamefont {Reuter}}, \bibinfo {author}
  {\bibfnamefont {A.}~\bibnamefont {Zappe}}, \bibinfo {author} {\bibfnamefont
  {F.~F.}\ \bibnamefont {de~Oliveira}}, \bibinfo {author} {\bibfnamefont
  {A.}~\bibnamefont {Denisenko}}, \bibinfo {author} {\bibfnamefont
  {H.}~\bibnamefont {Sumiya}}, \bibinfo {author} {\bibfnamefont
  {S.}~\bibnamefont {Onoda}}, \bibinfo {author} {\bibfnamefont
  {J.}~\bibnamefont {Isoya}},\ and\ \bibinfo {author} {\bibfnamefont
  {J.}~\bibnamefont {Wrachtrup}},\ }\bibfield  {title} {\bibinfo {title}
  {Nanoscale nuclear magnetic resonance with chemical resolution},\ }\href
  {https://doi.org/10.1126/science.aam8697} {\bibfield  {journal} {\bibinfo
  {journal} {Science}\ }\textbf {\bibinfo {volume} {357}},\ \bibinfo {pages}
  {67} (\bibinfo {year} {2017})}\BibitemShut {NoStop}%
\bibitem [{\citenamefont {Pfender}\ \emph
  {et~al.}(2017{\natexlab{b}})\citenamefont {Pfender}, \citenamefont {Aslam},
  \citenamefont {Sumiya}, \citenamefont {Onoda}, \citenamefont {Neumann},
  \citenamefont {Isoya}, \citenamefont {Meriles},\ and\ \citenamefont
  {Wrachtrup}}]{Pfender17Nonvolatile}%
  \BibitemOpen
  \bibfield  {author} {\bibinfo {author} {\bibfnamefont {M.}~\bibnamefont
  {Pfender}}, \bibinfo {author} {\bibfnamefont {N.}~\bibnamefont {Aslam}},
  \bibinfo {author} {\bibfnamefont {H.}~\bibnamefont {Sumiya}}, \bibinfo
  {author} {\bibfnamefont {S.}~\bibnamefont {Onoda}}, \bibinfo {author}
  {\bibfnamefont {P.}~\bibnamefont {Neumann}}, \bibinfo {author} {\bibfnamefont
  {J.}~\bibnamefont {Isoya}}, \bibinfo {author} {\bibfnamefont {C.~A.}\
  \bibnamefont {Meriles}},\ and\ \bibinfo {author} {\bibfnamefont
  {J.}~\bibnamefont {Wrachtrup}},\ }\bibfield  {title} {\bibinfo {title}
  {Nonvolatile nuclear spin memory enables sensor- unlimited nanoscale
  spectroscopy of small spin clusters},\ }\href@noop {} {\bibfield  {journal}
  {\bibinfo  {journal} {Nat. Comm.}\ }\textbf {\bibinfo {volume} {8}},\
  \bibinfo {pages} {834} (\bibinfo {year} {2017}{\natexlab{b}})}\BibitemShut
  {NoStop}%
\end{thebibliography}%

\end{document}